\documentclass[aps,pre,onecolumn,nofootinbib,superscriptaddress]{revtex4-2}

\usepackage[T1]{fontenc}
\usepackage[utf8]{inputenc}
\usepackage{lmodern}
\usepackage{amsmath,amssymb,amsthm,mathtools}
\usepackage{bm}
\usepackage{microtype}
\usepackage{hyperref}
\usepackage{enumitem}

\hypersetup{
  colorlinks=true,
  linkcolor=blue,
  citecolor=blue,
  urlcolor=blue,
  pdftitle={A Beltrami Mode in Extreme Fractional Hydrodynamics under Stochastic Forcing},
  pdfauthor={José I. H. López}
}

\newtheorem{proposition}{Proposition}

\newtheorem{remark}{Remark}
\newcommand{\T}{\mathbb T^3}
\newcommand{\R}{\mathbb R}
\newcommand{\E}{\mathbb E}

\newcommand{\ip}[2]{\left\langle #1,#2\right\rangle}
\newcommand{\norm}[1]{\left\|#1\right\|}

\begin{document}

\title{A Beltrami Mode in Extreme Fractional Hydrodynamics under Stochastic Forcing}

\author{Jos\'e I.~H. L\'opez}
\email{jihlpez@gmail.com}
\affiliation{Department of Mechanical Engineering, University of S\~ao Paulo (USP),
Av.\ Prof.\ Mello Moraes, 2231, 05508-030 S\~ao Paulo, SP, Brazil}
\date{\today}

\begin{abstract}
We construct an exact stochastic solution of the extreme-order endpoint of adaptive fractional Navier--Stokes (AFNS) hydrodynamics on the three-dimensional periodic torus. A helical Beltrami eigenfield reduces the nonlinear vector equation to a scalar Ornstein--Uhlenbeck process when the white-in-time forcing is aligned with that eigenfield. The resulting invariant Gaussian law yields explicit formulas for the stationary energy, its fluctuations, temporal correlations, circulation, and helicity. Under the same forcing, the corresponding classical Navier--Stokes mode has a stationary energy that diverges as the Reynolds number tends to infinity, while the Euler limit exhibits linear growth of the mean energy and has no invariant probability law. The construction isolates, without a cascade hypothesis, the statistical effect of the finite fractional dissipation retained at the extreme endpoint of AFNS.
\end{abstract}

\maketitle

\section{Introduction}

The high-Reynolds-number limit of viscous fluid equations is singular. In the classical incompressible Navier--Stokes equations, the coefficient multiplying the Laplacian tends to zero as the Reynolds number increases. For a fixed smooth mode, this removes the viscous relaxation and leads to the corresponding Euler dynamics. Turbulent flows, however, motivate a more delicate question: can an effective dissipative mechanism remain relevant in a high-Reynolds-number description even when the molecular-viscosity coefficient vanishes?

Adaptive fractional Navier--Stokes (AFNS) addresses this question by allowing the order of the dissipative operator to depend on the local Reynolds regime. In the formulation introduced in the foundational AFNS preprint \cite{Lopez2026AFNS}, the operator interpolates between the classical Laplacian at the regular end ($s\to1$) and the fractional Laplacian $(-\Delta)^{s}$ with $s=1/3$, that is, an operator of differential order $2/3$, in the extreme three-dimensional regime. The latter endpoint retains a finite coefficient set by the critical Reynolds number. The present paper does not attempt to rederive the adaptive closure or to establish the full variable-order model. Instead, it isolates the extreme endpoint and asks a narrower question: what does its residual dissipation do in a setting where the nonlinear dynamics can be treated exactly?

Beltrami fields provide a natural setting for this purpose. They are divergence-free eigenfields of curl, and their velocity and vorticity are parallel \cite{ArnoldKhesin1998,MajdaBertozzi2002}. The Lamb identity then makes the convective term a pure gradient. For a single Beltrami eigenfield, the nonlinear term can therefore be absorbed into the pressure, while any spectral function of the Laplacian acts diagonally. This structure gives an exact invariant one-dimensional family of velocity fields. It allows the deterministic fractional equation, the classical Navier--Stokes equation, and the Euler equation to be compared on precisely the same spatial profile. Exact periodic Navier--Stokes solutions built from non-interacting Fourier modes are classified in Ref.~\cite{Bogoyavlenskii2003}.

We then introduce a white-in-time random force aligned with that Beltrami field. This choice is deliberate: it injects energy into one known spectral mode and preserves the invariant family. The resulting infinite-dimensional stochastic fluid equation reduces exactly to a one-dimensional Ornstein--Uhlenbeck process. Consequently, its invariant law and all of its basic energy statistics can be computed without closure assumptions, spectral truncations, or numerical approximations.

The comparison with Euler is especially transparent. Under the same additive forcing, the extreme AFNS mode has a finite relaxation rate and reaches a stationary Gaussian distribution. The corresponding Euler amplitude is Brownian motion, so its mean energy grows linearly and no invariant probability measure exists. At finite Reynolds number the classical Navier--Stokes mode is also Ornstein--Uhlenbeck, but its stationary energy grows proportionally to the Reynolds number; its invariant measures therefore fail to remain tight in the infinite-Reynolds-number limit. The AFNS endpoint behaves differently because its modal relaxation rate remains finite.

The scope of the result must be stated carefully. The solution constructed below is an exact stochastic Beltrami flow, not a generic turbulent solution with a developed cascade. Energy is injected and dissipated directly in the same spectral mode. The value of the construction is that it isolates one precise consequence of the extreme AFNS operator: the ability to regulate the energy of a continuously forced mode at infinite macroscopic Reynolds number. It supplies a controlled analytical test case and a starting point for later stability questions involving perturbations away from the Beltrami family.

The paper is organized as follows. Section II fixes the periodic setting and records the Beltrami identities. Section III derives the exact deterministic modal reductions for classical Navier--Stokes, extreme AFNS, and Euler. Section IV introduces the stochastic forcing and proves the exact Ornstein--Uhlenbeck reduction. Section V develops the stationary and transient statistics. Section VI compares the three models under identical forcing. Section VII discusses the physical interpretation and the limitations of the construction, and Section VIII concludes.

\section{Periodic setting and Beltrami geometry}

We work on the flat torus
\[
  \T=(\R/2\pi\mathbb Z)^3,
\]
with normalized spatial inner product
\[
  \ip{v}{w}
  =\frac{1}{(2\pi)^3}\int_{\T}v(x)\cdot w(x)\,dx,
  \qquad
  \norm{v}_2^2=\ip{v}{v}.
\]
All vector fields are periodic, divergence-free, and have zero spatial mean.

A Beltrami field is a nonzero vector field $U$ satisfying
\begin{equation}
  \nabla\cdot U=0,
  \qquad
  \nabla\times U=\lambda U,
  \qquad \lambda>0.
  \label{eq:beltrami}
\end{equation}
Taking the curl of the second relation and using incompressibility gives
\begin{equation}
  -\Delta U=\lambda^2 U.
  \label{eq:laplace-eigen}
\end{equation}
Thus $U$ is simultaneously a curl eigenfield and a Laplacian eigenfield.

For definiteness, one may choose the explicit periodic field
\begin{equation}
  U_\lambda(x)
  =
  \begin{pmatrix}
    \sin(\lambda x_3)\\
    \cos(\lambda x_3)\\
    0
  \end{pmatrix},
  \qquad \lambda\in\mathbb N.
  \label{eq:explicit-U}
\end{equation}
It satisfies
\[
  \nabla\cdot U_\lambda=0,\qquad
  \nabla\times U_\lambda=\lambda U_\lambda,\qquad
  -\Delta U_\lambda=\lambda^2U_\lambda,
\]
and has unit norm in the normalized inner product above. The analysis below also applies to any normalized Beltrami eigenfield satisfying Eq.~(\ref{eq:beltrami}).

The Lamb identity \cite{MajdaBertozzi2002} reads
\begin{equation}
  (U\cdot\nabla)U
  =\nabla\left(\frac{|U|^2}{2}\right)
   -U\times(\nabla\times U).
  \label{eq:lamb}
\end{equation}
For a Beltrami field the cross product vanishes, and hence
\begin{equation}
  (U\cdot\nabla)U
  =\nabla\left(\frac{|U|^2}{2}\right).
  \label{eq:beltrami-nonlinearity}
\end{equation}
Consequently, for any scalar amplitude $a(t)$,
\begin{equation}
  (aU\cdot\nabla)(aU)
  =\nabla\left(\frac{a(t)^2|U|^2}{2}\right).
  \label{eq:scaled-nonlinearity}
\end{equation}
This term is exactly canceled by the pressure choice
\[
  p(x,t)=-\frac{a(t)^2}{2}|U(x)|^2
\]
(up to an arbitrary function of time). Equivalently, its Leray projection onto divergence-free vector fields is zero.

The fractional Laplacian is defined spectrally on the torus (for the relation between spectral and integral definitions on $\R^n$, see Ref.~\cite{DiNezza2012}). From Eq.~(\ref{eq:laplace-eigen}),
\begin{equation}
  \Lambda^{2s}U=(-\Delta)^sU=\lambda^{2s}U,
  \qquad s>0,
  \label{eq:fractional-eigen}
\end{equation}
where $\Lambda=(-\Delta)^{1/2}$. In particular,
\begin{equation}
  \Lambda^{2/3}U=\lambda^{2/3}U.
  \label{eq:two-thirds-eigen}
\end{equation}

Equations~(\ref{eq:beltrami-nonlinearity}) and (\ref{eq:fractional-eigen}) are the two structural facts used throughout: the nonlinearity contributes only to the pressure, and the dissipative operator acts as a scalar on the chosen field.

\section{Deterministic modal reduction}

\subsection{Classical Navier--Stokes and its Euler limit}

In dimensionless variables, the incompressible Navier--Stokes equations on $\T$ are
\begin{equation}
  \partial_t u+(u\cdot\nabla)u+\nabla p
  -\frac{1}{Re}\Delta u=0,
  \qquad \nabla\cdot u=0.
  \label{eq:NS}
\end{equation}
Substitute
\begin{equation}
  u(x,t)=a(t)U(x).
  \label{eq:ansatz}
\end{equation}
The convective term is absorbed into the pressure by Eq.~(\ref{eq:scaled-nonlinearity}), and the Laplacian acts diagonally by Eq.~(\ref{eq:laplace-eigen}). The vector equation therefore reduces to
\begin{equation}
  \dot a(t)=-\kappa_{\mathrm{NS}}a(t),
  \qquad
  \kappa_{\mathrm{NS}}=\frac{\lambda^2}{Re}.
  \label{eq:NS-amplitude}
\end{equation}
The exact solution is
\begin{equation}
  u_{\mathrm{NS}}(x,t)
  =a_0e^{-\lambda^2t/Re}U(x).
  \label{eq:NS-exact}
\end{equation}

For fixed $t$ and fixed smooth initial mode, the limit $Re\to\infty$ gives
\[
  u_{\mathrm{NS}}(x,t)\longrightarrow a_0U(x).
\]
This is the corresponding stationary Euler Beltrami solution. The convergence here concerns this exact fixed-profile mode; it does not address singular high-Reynolds-number limits in which gradients or spectral content grow with $Re$.

\subsection{The extreme AFNS endpoint}

The AFNS framework promotes the fractional order of the dissipative operator to a Reynolds-dependent quantity. In the notation of the foundational formulation \cite{Lopez2026AFNS}, the order approaches $s=1/3$ in the extreme high-local-Reynolds-number regime, while the dissipative amplitude approaches $Re_c^{-1}$. We study that endpoint as a constant-coefficient reduced model:
\begin{equation}
  \partial_t u+(u\cdot\nabla)u+\nabla p
  +\frac{1}{Re_c}\Lambda^{2/3}u=0,
  \qquad \nabla\cdot u=0.
  \label{eq:AFNS-extreme}
\end{equation}
This equation is the extreme endpoint of AFNS, not the full variable-order equation at finite local Reynolds number.

Using the same ansatz (\ref{eq:ansatz}) and the same Beltrami field, Eq.~(\ref{eq:AFNS-extreme}) reduces exactly to
\begin{equation}
  \dot a(t)=-\kappa_Aa(t),
  \qquad
  \kappa_A=\frac{\lambda^{2/3}}{Re_c}.
  \label{eq:AFNS-amplitude}
\end{equation}
Thus
\begin{equation}
  u_A(x,t)
  =a_0\exp\left(-\frac{\lambda^{2/3}}{Re_c}t\right)U(x).
  \label{eq:AFNS-exact}
\end{equation}

The distinction from classical Navier--Stokes is spectral. For the same mode,
\[
  \kappa_{\mathrm{NS}}=\frac{\lambda^2}{Re},
  \qquad
  \kappa_A=\frac{\lambda^{2/3}}{Re_c}.
\]
The classical modal relaxation vanishes as $Re\to\infty$, whereas the extreme AFNS modal relaxation remains positive for fixed $\lambda$ and $Re_c$.

\begin{remark}
The reduction is exact because the operator order and coefficient have been fixed at their extreme endpoint. If the local Reynolds number is recalculated dynamically from the stochastic velocity amplitude, then $s$ and the dissipative coefficient may become time-dependent. The scalar equation need not remain Ornstein--Uhlenbeck in that fully coupled problem. That extension is not assumed here.
\end{remark}

\section{Stochastic forcing and exact reduction}

We now force the extreme AFNS equation with a white-in-time random force aligned with the Beltrami eigenfield:
\begin{equation}
  du+\left[(u\cdot\nabla)u+\nabla p
  +\frac{1}{Re_c}\Lambda^{2/3}u\right]dt
  =\sigma U\,dW_t,
  \qquad \nabla\cdot u=0.
  \label{eq:stoch-AFNS}
\end{equation}
Here $W_t$ is a standard real Brownian motion and $\sigma>0$ is the forcing amplitude. The stochastic differential is interpreted in the It\^o sense. Since the noise is additive, the It\^o and Stratonovich formulations of the velocity equation coincide; the It\^o correction appears when nonlinear observables such as the energy are computed.

\begin{proposition}[Exact stochastic Beltrami solution]
Let $U$ be a normalized Beltrami eigenfield satisfying Eqs.~(\ref{eq:beltrami})--(\ref{eq:laplace-eigen}). If the initial condition is $u(x,0)=a_0U(x)$ and the forcing in Eq.~(\ref{eq:stoch-AFNS}) is $\sigma U\,dW_t$, then
\begin{equation}
  u(x,t)=a_tU(x)
  \label{eq:stoch-ansatz}
\end{equation}
is an exact solution, where the scalar amplitude solves
\begin{equation}
  da_t=-\kappa_Aa_t\,dt+\sigma\,dW_t,
  \qquad \kappa_A=\frac{\lambda^{2/3}}{Re_c}.
  \label{eq:OU}
\end{equation}
\end{proposition}

\begin{proof}
Substitute $u=a_tU$ into Eq.~(\ref{eq:stoch-AFNS}). The nonlinear term is the gradient in Eq.~(\ref{eq:scaled-nonlinearity}) and is canceled by the pressure. The fractional dissipative term is
\[
  \frac{1}{Re_c}\Lambda^{2/3}(a_tU)\,dt
  =\frac{\lambda^{2/3}}{Re_c}a_tU\,dt.
\]
The remaining vector-valued stochastic equation is
\[
  U\left(da_t+\kappa_Aa_t\,dt-\sigma\,dW_t\right)=0.
\]
Since $U$ is nonzero, the scalar factor must vanish, giving Eq.~(\ref{eq:OU}).
\end{proof}

The solution of Eq.~(\ref{eq:OU}) is, by standard stochastic calculus \cite{Gardiner2009},
\begin{equation}
  a_t=a_0e^{-\kappa_At}
  +\sigma\int_0^t e^{-\kappa_A(t-r)}\,dW_r.
  \label{eq:OU-solution}
\end{equation}
This proves that the infinite-dimensional stochastic fluid equation possesses an invariant one-dimensional family of solutions. On that family, the full dynamics is exactly the Ornstein--Uhlenbeck process.

\section{Statistical properties}

\subsection{Amplitude law and invariant measure}

For deterministic $a_0$, the amplitude is Gaussian at every positive time, with
\begin{align}
  \E[a_t]&=a_0e^{-\kappa_At},\\
  \operatorname{Var}(a_t)
  &=\frac{\sigma^2}{2\kappa_A}
    \left(1-e^{-2\kappa_At}\right).
  \label{eq:mean-var}
\end{align}
As $t\to\infty$, the transient mean vanishes and the variance converges to $\sigma^2/(2\kappa_A)$. The invariant density is
\begin{equation}
  \rho_A(a)
  =\sqrt{\frac{\kappa_A}{\pi\sigma^2}}
    \exp\left(-\frac{\kappa_Aa^2}{\sigma^2}\right),
  \qquad a\in\R.
  \label{eq:invariant-density}
\end{equation}
Equivalently,
\begin{equation}
  a_\infty\sim\mathcal N\left(0,\frac{\sigma^2}{2\kappa_A}\right).
  \label{eq:stationary-gaussian}
\end{equation}
This is the unique invariant probability measure for the scalar Ornstein--Uhlenbeck process \cite{DaPratoZabczyk1996}. It is also the invariant law induced on the one-dimensional velocity subspace $\{aU:a\in\R\}$. Since the noise is nondegenerate on this subspace and the drift is strictly dissipative, the reduced process is ergodic and exponentially mixing.

In stationarity, the two-time amplitude correlation is
\begin{equation}
  \E[a_ta_{t+\tau}]
  =\frac{\sigma^2}{2\kappa_A}e^{-\kappa_A|\tau|}.
  \label{eq:amplitude-correlation}
\end{equation}
The corresponding correlation time is
\begin{equation}
  \tau_A=\kappa_A^{-1}
  =\frac{Re_c}{\lambda^{2/3}}.
  \label{eq:correlation-time}
\end{equation}

\subsection{Energy balance and stationary energy}

Because $\norm{U}_2=1$, the kinetic energy is
\begin{equation}
  E_A(t)=\frac12\norm{u(t)}_2^2=\frac12a_t^2.
  \label{eq:energy}
\end{equation}
Applying It\^o's formula to $E_A=a_t^2/2$ gives
\begin{equation}
  dE_A
  =-\kappa_Aa_t^2\,dt
   +\sigma a_t\,dW_t
   +\frac{\sigma^2}{2}\,dt.
  \label{eq:ito-energy}
\end{equation}
The three terms have distinct meanings. The first is the instantaneous fractional dissipation, the second is the fluctuating work of the random force, and the third is the positive It\^o contribution responsible for the mean energy injection.

Taking expectations yields the closed energy equation
\begin{equation}
  \frac{d}{dt}\E[E_A]
  =-2\kappa_A\E[E_A]+\frac{\sigma^2}{2}.
  \label{eq:mean-energy-ode}
\end{equation}
Its solution is
\begin{equation}
  \E[E_A(t)]
  =E_A(0)e^{-2\kappa_At}
   +\frac{\sigma^2}{4\kappa_A}
    \left(1-e^{-2\kappa_At}\right).
  \label{eq:mean-energy}
\end{equation}
Therefore,
\begin{equation}
  \boxed{
  \E[E_A]_{\mathrm{stat}}
  =\frac{\sigma^2}{4\kappa_A}
  =\frac{\sigma^2Re_c}{4\lambda^{2/3}}.
  }
  \label{eq:stationary-energy}
\end{equation}

The mean energy injection rate is
\begin{equation}
  \mathcal I=\frac{\sigma^2}{2},
  \label{eq:injection}
\end{equation}
whereas the instantaneous dissipation rate is
\begin{equation}
  \mathcal D_A(t)
  =\frac{1}{Re_c}\ip{u}{\Lambda^{2/3}u}
  =\kappa_Aa_t^2
  =2\kappa_AE_A(t).
  \label{eq:dissipation}
\end{equation}
At stationarity,
\begin{equation}
  \boxed{
  \E[\mathcal D_A]_{\mathrm{stat}}
  =\mathcal I=\frac{\sigma^2}{2}.
  }
  \label{eq:balance}
\end{equation}
Thus energy is injected and dissipated in the same Beltrami mode. No intermodal transfer or inertial-range cascade is required for this exact balance.

\subsection{The stationary energy distribution}

The stationary amplitude is centered Gaussian. Since $E_A=a^2/2$, the energy has a scaled chi-square distribution with one degree of freedom. Its density is
\begin{equation}
  p_{E_A}(e)
  =\sqrt{\frac{2\kappa_A}{\pi\sigma^2e}}
    \exp\left(-\frac{2\kappa_Ae}{\sigma^2}\right),
  \qquad e>0.
  \label{eq:energy-density}
\end{equation}
In addition to its mean, the energy variance is
\begin{equation}
  \operatorname{Var}(E_A)
  =\frac{\sigma^4}{8\kappa_A^2}
  =\frac{\sigma^4Re_c^2}{8\lambda^{4/3}}.
  \label{eq:energy-variance}
\end{equation}
The stationary energy autocovariance is
\begin{equation}
  \operatorname{Cov}(E_A(t),E_A(t+\tau))
  =\frac{\sigma^4}{8\kappa_A^2}
    e^{-2\kappa_A|\tau|}.
  \label{eq:energy-correlation}
\end{equation}
The energy therefore decorrelates twice as fast as the amplitude, as expected for the square of a stationary Gaussian process.

\subsection{Circulation, helicity, and enstrophy}

Let $C$ be a fixed oriented closed curve on the torus. Its circulation is
\begin{equation}
  \Gamma_C(t)=\oint_Cu(x,t)\cdot dx
  =a_t\Gamma_U(C),
  \qquad
  \Gamma_U(C)=\oint_CU\cdot dx.
  \label{eq:circulation}
\end{equation}
It is therefore a centered Gaussian observable in stationarity, with
\begin{align}
  \E[\Gamma_C]&=0,\\
  \operatorname{Var}(\Gamma_C)
  &=\Gamma_U(C)^2\frac{\sigma^2}{2\kappa_A},\\
  \E[\Gamma_C(t)\Gamma_C(t+\tau)]
  &=\Gamma_U(C)^2\frac{\sigma^2}{2\kappa_A}
    e^{-\kappa_A|\tau|}.
\end{align}

The helicity is
\[
  H_A(t)=\int_{\T}u\cdot(\nabla\times u)\,
  \frac{dx}{(2\pi)^3}.
\]
Using $\nabla\times U=\lambda U$ and $\norm{U}_2=1$,
\begin{equation}
  H_A(t)=\lambda a_t^2=2\lambda E_A(t).
  \label{eq:helicity}
\end{equation}
This relation holds pathwise, not merely after averaging. Hence
\begin{equation}
  \E[H_A]_{\mathrm{stat}}
  =\frac{\lambda\sigma^2}{2\kappa_A}
  =\frac{\sigma^2Re_c\,\lambda^{1/3}}{2},
  \qquad
  \operatorname{Var}(H_A)
  =4\lambda^2\operatorname{Var}(E_A).
  \label{eq:helicity-stats}
\end{equation}
Similarly, the enstrophy $\mathcal Z_A=\frac12\norm{\nabla\times u}_2^2$ satisfies
\begin{equation}
  \mathcal Z_A(t)=\lambda^2E_A(t).
  \label{eq:enstrophy}
\end{equation}
All these observables inherit their statistics from the same scalar stationary process.

\section{Comparison with stochastically forced Navier--Stokes and Euler}

To isolate the role of the dissipative operator, we impose exactly the same noise $\sigma U\,dW_t$ on the classical Navier--Stokes and Euler equations. The initial condition and spatial mode are also identical.

\subsection{Finite-Reynolds-number Navier--Stokes}

For the forced classical equation,
\[
  du+\left[(u\cdot\nabla)u+\nabla p
  -\frac{1}{Re}\Delta u\right]dt
  =\sigma U\,dW_t,
\]
the Beltrami ansatz gives
\begin{equation}
  da_{\mathrm{NS}}
  =-\frac{\lambda^2}{Re}a_{\mathrm{NS}}\,dt
   +\sigma\,dW_t.
  \label{eq:OU-NS}
\end{equation}
At every finite $Re$, this is an Ornstein--Uhlenbeck process. Its invariant amplitude variance and mean energy are
\begin{equation}
  \operatorname{Var}(a_{\mathrm{NS}})_{\mathrm{stat}}
  =\frac{\sigma^2Re}{2\lambda^2},
  \qquad
  \E[E_{\mathrm{NS}}]_{\mathrm{stat}}
  =\frac{\sigma^2Re}{4\lambda^2}.
  \label{eq:NS-stationary-energy}
\end{equation}
Thus the classical equation does have a stationary law for every finite Reynolds number in this one-mode problem. However, its stationary energy diverges linearly with $Re$. The family of stationary Gaussian measures is not tight as $Re\to\infty$.

\subsection{Euler limit}

At infinite Reynolds number, the modal damping in Eq.~(\ref{eq:OU-NS}) vanishes. The forced Euler amplitude is
\begin{equation}
  da_E=\sigma\,dW_t,
  \qquad
  a_E(t)=a_0+\sigma W_t.
  \label{eq:Euler-BM}
\end{equation}
It follows that
\begin{equation}
  \boxed{
  \E[E_E(t)]
  =E_E(0)+\frac{\sigma^2}{2}t.
  }
  \label{eq:Euler-energy}
\end{equation}
The energy grows linearly in expectation, and Brownian motion on $\R$ has no invariant probability measure.

\subsection{The comparison}

The three modal relaxation rates are
\begin{equation}
  \kappa_{\mathrm{NS}}=\frac{\lambda^2}{Re},
  \qquad
  \kappa_A=\frac{\lambda^{2/3}}{Re_c},
  \qquad
  \kappa_E=0.
  \label{eq:three-rates}
\end{equation}
For fixed $\lambda$ and $Re_c$, the AFNS rate remains positive as $Re\to\infty$, whereas the classical Navier--Stokes rate tends to zero. This produces three distinct statistical regimes:
\begin{enumerate}[leftmargin=2em]
  \item \textit{Finite-$Re$ Navier--Stokes:} a stationary Gaussian amplitude exists, but its variance grows proportionally to $Re$.
  \item \textit{Euler:} the amplitude is Brownian, the mean energy grows linearly in time, and no invariant probability measure exists.
  \item \textit{Extreme AFNS:} the amplitude is Ornstein--Uhlenbeck with a nonzero limiting relaxation rate, and the energy has a finite stationary distribution.
\end{enumerate}

The difference is not caused by a change in forcing, geometry, or initial data. It is entirely due to the dissipative operator retained by the extreme AFNS endpoint.

A useful way to express the contrast is through the stationary energy:
\begin{equation}
  \E[E_{\mathrm{NS}}]_{\mathrm{stat}}
  =\frac{\sigma^2Re}{4\lambda^2}
  \longrightarrow\infty,
  \qquad
  \E[E_A]_{\mathrm{stat}}
  =\frac{\sigma^2Re_c}{4\lambda^{2/3}}<\infty.
  \label{eq:stationary-contrast}
\end{equation}
The AFNS stationary law therefore remains well-defined in the extreme macroscopic-Reynolds-number regime considered here, while the classical stationary laws escape to arbitrarily large amplitudes. More precisely, $\E[E_{\mathrm{NS}}]_{\mathrm{stat}}/\E[E_A]_{\mathrm{stat}}=(Re/Re_c)\,\lambda^{-4/3}$, so the AFNS endpoint is the more strongly dissipative one exactly when $Re>Re_c\lambda^{4/3}$; the contrast above refers to $Re\to\infty$ at fixed $\lambda$.

\section{Discussion and limitations}

The construction gives an exact example of statistical energy regulation by fractional dissipation at an extreme Reynolds endpoint. Its central mechanism is simple. The noise injects energy directly into a Beltrami eigenmode. The fractional operator acts diagonally on that same mode and removes energy at a rate proportional to its instantaneous energy. The balance produces an Ornstein--Uhlenbeck process and an explicit invariant distribution.

Several distinctions are important for interpreting this result.

First, this is not a cascade calculation. The forcing and dissipation act on the same eigenmode, and the Beltrami structure removes the nonlinear transfer term after projection onto divergence-free fields. The example therefore does not establish a Kolmogorov spectrum, an inertial range, or a dissipative anomaly generated by nonlinear transfer. It isolates the direct modal effect of the extreme fractional operator.

Second, the invariant probability measure is established for the scalar reduced dynamics, or equivalently for the dynamics restricted to the invariant line spanned by $U$. The argument does not prove existence, uniqueness, or ergodicity of an invariant measure for the full three-dimensional AFNS stochastic partial differential equation with general initial data and general noise; for the classical stochastic three-dimensional Navier--Stokes equations, see Ref.~\cite{DaPratoDebussche2003}.

Third, the extreme operator has been frozen at $s=1/3$ and coefficient $Re_c^{-1}$. This is the asymptotic endpoint motivated by the adaptive AFNS formulation \cite{Lopez2026AFNS}. If the local Reynolds number is allowed to fluctuate with the stochastic velocity, then the order and coefficient may themselves become random and time-dependent. The resulting amplitude equation would generally no longer be the linear Ornstein--Uhlenbeck equation derived here. Establishing that self-consistent adaptive stochastic problem is a separate task.

Fourth, the comparison with Euler is specific and controlled. It shows that Euler cannot balance this particular additive white-noise injection in the single Beltrami mode. It does not imply that every stochastically forced Euler system lacks stationary statistical states; the conclusion depends on the forcing, the invariant mode, and the absence of a dissipative or other regulating mechanism in the reduced equation.

These limitations do not weaken the exact statement. They define its domain of validity and make the next mathematical question precise: whether the stationary Beltrami statistics persist, approximately or exactly, under perturbations that leave the Beltrami eigenspace.

\section{Conclusions}

We have constructed an exact stochastic solution of the extreme fractional AFNS equation on the three-dimensional periodic torus. The construction uses a Beltrami eigenfield, for which the convective term is a pressure gradient and the fractional Laplacian acts by multiplication with the corresponding spectral eigenvalue. A white-in-time force aligned with that field preserves the invariant Beltrami family and reduces the full vector equation to a scalar Ornstein--Uhlenbeck process.

This reduction yields the complete stationary statistics of the mode. In particular, the energy has a finite stationary mean proportional to $Re_c\lambda^{-2/3}$, a finite variance, and an explicitly known probability density. The mean energy injection is exactly balanced by the mean fractional dissipation. Circulation, helicity, and enstrophy are also explicit observables of the same stationary process.

Under the same forcing and on the same spatial mode, classical Navier--Stokes has a stationary distribution at finite Reynolds number, but its stationary energy grows proportionally to $Re$ and therefore diverges in the high-Reynolds-number limit. The corresponding Euler amplitude is Brownian motion, with linearly growing mean energy and no invariant probability law. The extreme AFNS endpoint differs because its modal relaxation rate remains finite.

The result is deliberately narrower than a theory of turbulence: it contains no cascade and makes no claim about generic three-dimensional stochastic flows. Its significance is that it provides an exact, analytically tractable test of the finite dissipative capacity retained by AFNS at its extreme endpoint. It identifies a concrete statistical distinction between that endpoint and the Euler limit, with identical forcing and geometry. The natural next step is to test the persistence of this regulation when the flow is perturbed away from the Beltrami family and when the adaptive dependence of the operator on the local Reynolds number is restored.

\end{document}